\documentclass[aps,prd,showpacs,nofootinbib,superscriptaddress,preprint,tightenlines]{revtex4}
\usepackage{graphicx}
\usepackage{amsbsy}

\usepackage{amsfonts}

\usepackage{amssymb}

\usepackage{amsmath}
\newcommand{\hateq}{\widehat{=}}
\def\be{\nopagebreak[3]\begin{equation}}
\def\ee{\end{equation}}
\def\ba{\nopagebreak[3]\begin{eqnarray}}
\def\ea{\end{eqnarray}}

\def\d{{\rm d}}

\def\a{\alpha}

\def\f{\frac}
\def\G{\Gamma}
\def\lp{\ell_{\rm Pl}}

\def\H{{\cal H}}

\def\G{{\Gamma}}   \def\H{{\cal H}} 
\def\g{{\gamma}}  \def\a{{\alpha}} 
\def\U(1){{\rm U(1)}}   \def\S{{\cal S}}

\def\d{{\rm d}}

\def\bar{\overline}

\def\={\hateq}
\def\SU(2){\rm SU(2)}
\def\U(1){\rm U(1)}
\def\ub{\underbar}
\def\half{{\textstyle{1\over2}}}

\newcommand{\teta}{\rlap{\lower2ex\hbox{$\,\tilde{}$}}\eta{}}

\newcommand{\tE}{\tilde{E}}

\begin{document}
\preprint{\vbox{\baselineskip=12pt \rightline{IGPG-07/3-1}
\rightline{gr-qc/0703116} }}
\title{Loop quantum gravity and
Planck-size black hole entropy}
\author{Alejandro Corichi}
\email{corichi@matmor.unam.mx} \affiliation{Instituto de
Matem\'aticas, Unidad Morelia, Universidad Nacional Aut\'onoma de
M\'exico, UNAM-Campus Morelia, A. Postal 61-3, Morelia,
Michoac\'an 58090, Mexico}
\affiliation{Instituto de Ciencias
Nucleares,
 Universidad Nacional Aut\'onoma de M\'exico,\\
A. Postal 70-543, M\'exico D.F. 04510, Mexico}
\affiliation{Institute for Gravitational Physics and Geometry,
Physics Department, Pennsylvania State University, University Park
PA 16802, USA}

\author{Jacobo D\'\i az-Polo}
\email{Jacobo.Diaz@uv.es}\affiliation{Departamento de Astronom\'\i
a y Astrof\'\i sica, Universidad de Valencia, Burjassot-46100,
Valencia, Spain}

\author{Enrique Fern\'andez-Borja}\email{Enrique.Fernandez@uv.es}
\affiliation{Departamento de F\'\i sica Te\'orica and IFIC, Centro
Mixto Universidad de Valencia-CSIC. Universidad de Valencia,
Burjassot-46100, Valencia, Spain}


\begin{abstract}
The Loop Quantum Gravity (LQG) program  is briefly reviewed and
one of its main applications, namely the counting of black hole
entropy within the framework is considered. In particular, recent
results for Planck size black holes are reviewed. These results
are consistent with an asymptotic linear relation (that fixes
uniquely a free parameter of the theory) and a logarithmic
correction with a coefficient equal to $-1/2$. The account is
tailored as an introduction to the subject for non-experts.
\end{abstract}

 \maketitle

\section{Introduction}

\noindent {\it Loop Quantum Gravity} (LQG) has become in the past
years a mayor player as a candidate for a quantum theory of
gravity \cite{lqg,AL:review}. On the one hand it has matured into
a serious contender together with other approaches such as {\it
String/M Theory}, but on the other is it not as well understood,
neither properly credited as a real physical theory of quantum
gravity. The purpose of this contribution is twofold. First, we
would like to provide a starting point for those interested in
learning the basics of the theory and to provide at the same time
an introduction to one of its application, namely the treatment of
black holes and its entropy counting.

Loop quantum gravity is based and two basic principles, namely the
general principles of quantum theory and the lessons from general
relativity: that physics is diffeomorphism invariant. This means
that the field describing the gravitational interaction, and the
geometry of spacetime is fully dynamical and interacting with the
rest of the fields present; when one considers the description of
the full gravity-matter system, this better be a background
independent one. The fact that LQG is based in general principles
of quantum mechanics means only that one is looking for a
description based on the language of Quantum Mechanics: states are
elements on a Hilbert space (well defined, one expects),
observables will be Hermitian operators thereon, etc. This does
not mean that one should use {\it all} that is already known about
quantizing fields. Quite on the contrary, the tools needed to
construct a background independent quantization (certainly not
like the quantization we know), are rather new.

Another important feature of LQG is that it is the most serious
attempt to perform a full {\it non-perturbative} quantization of
the gravitational field by itself. It is an attempt to answer the
following question: can we quantize the gravitational degrees of
freedom without considering matter on the first place? Since LQG
aims at being a physical theory, which means it better be
falsifiable, one expects to answer that question unambiguously,
whenever one has the theory fully developed. This is one of the
main present challenges of the theory, namely to produce
predictions that can be tested experimentally. On the other hand,
one can consider situations in which the full knowledge of the
quantum theory is not needed in order to describe a particular
physical situation. This is precisely the case of black hole
horizons and their entropy.

Black hole entropy is one of the most intriguing constructs of
modern theoretical physics. On the one hand, it has a
correspondence with the black hole horizon area through the laws
of (classical) black hole mechanics \cite{BCH}. On the other hand
it is assumed to have a quantum statistical origin given that the
proper identification between entropy and area $S=A/4 \,\ell^2_p$
came only after an analysis of {\it quantum} fields on a fixed
background \cite{BH}. One of the most crucial test that a
candidate quantum theory of gravity must pass then is to provide a
mechanism to account for the microscopic degrees of freedom of
black holes. It is not unfair to say that at the moment we have
only two candidates for quantum gravity that have offered such an
explanation: string/brane theory \cite{vafa} and loop quantum
gravity \cite{ABCK}. The LQG formalism uses as starting point
isolated horizon (IH) boundary conditions at the classical level,
where the interior of the BH is excluded from the region under
consideration. In this sense, the description is somewhat
effective, since part of the information about the interior is
encoded in the boundary conditions at the horizon that in the
quantum theory get promoted to a condition that horizon states
must satisfy. There is also an important issue regarding this
formalism. Loop quantum gravity possesses a one parameter family
of inequivalent representations of the classical theory labelled
by a real number $\gamma$, the so called Barbero-Immirzi (BI)
parameter  (it is the  analogue of the $\theta$ ambiguity in QCD
\cite{BI}). It turns out that the BH entropy calculation provides
a linear relation between entropy and area for very large black
holes (in Planck units) as,
$$S=\lambda\,A(\gamma),$$
 where the parameter
$\lambda$ is independent of $\gamma$ and depends only in the
counting. We have put the $\gamma$ dependence in the Area, since
the parameter appears explicitly in the area spectrum. The
strategy that has been followed within the LQG community is to
regard the Bekenstein-Hawking entropy $S=A/4$ as a requirement
that large black holes  should satisfy. This fixes uniquely the
value of $\gamma=\gamma_0$ once and for all, by looking at the
asymptotic behavior, provided that one has the `correct counting'
that provides the right value for $\gamma_0$. The analytic
counting has provided also an expression for the first correction
term that turns out to be logarithmic \cite{meiss,GM}.

The purpose of the present article is to provide an introduction
to the main ideas behind the loop quantum gravity program and to
one of its main applications, namely the computation of black hole
entropy. In this regard, we shall describe some recent results
that have been performed for \emph{Planck size} black holes and
that complement in a precise way the analytical computations. In
particular, as we shall show, even when the black holes considered
are outside the original domain of applicability of the formalism,
one can still learn from these considerations.

The structure of the paper is as follows. In Sec.~\ref{sec:2} we
present some preliminaries, such as the standard geometrodynamical
variables for canonical gravity, the passage to connections and
triads and the choice of classical observables to be quantized. In
Sec.~\ref{sec:3} we describe the loop quantum geometry formalism,
including some relevant geometric operators. Sec.~\ref{sec:4} is
devoted to the formalism of quantum isolated horizons. We recall
the classical formalism and the basic steps to the quantization of
the horizon theory. State counting and black hole entropy is the
subject of Sec.~\ref{sec:5}. We and with a discussion in
Sec.~\ref{sec:6}.

\section{Preliminaries}
\label{sec:2}

\subsection{Geometrodynamics}

\noindent The first step is to introduce the basic classical
variables of the theory. Since the theory is described by a
Hamiltonian formalism, this means that the 4-dim spacetime $M$ of
the form $M=\Sigma\times\mathbb{R}$, where $\Sigma$ is a
3-dimensional manifold. The first thing to do is to start with the
geometrodynamical phase space $\G_{\rm g}$ of Riemannian metrics
$q_{ab}$ on $\Sigma$ and their canonical momenta
$\tilde{\pi}^{ab}$ (related to the extrinsic curvature $K_{ab}$ of
$\Sigma$ into $M$ by
$\tilde{\pi}^{ab}=\sqrt{q}\,(K^{ab}-\frac{1}{2}q^{ab}\,K)$, with
$q={\rm det}(q_{ab})$ and $K=q^{ab}K_{ab}$). Recall that they
satisfy,
\be
\{\tilde{\pi}^{ab}(x),q_{cd}(y)\}=2\kappa\,\delta^{a}_{(c}\,\delta^b_{d)}
\,\delta^3(x,y)\quad
;\quad
\{q_{ab}(x),q_{cd}(y)\}=\{\tilde{\pi}^{ab}(x),\tilde{\pi}^{cd}(y)\}=0
\ee
General Relativity in these geometrodynamical variables is a
theory with constraints, which means that the canonical variables
$(q_{ab},\tilde{\pi}^{ab})$ do not take arbitrary values but must
satisfy four constraints: \be {\cal
H}^b=D_a\,(\tilde{\pi}^{ab})\approx 0 \qquad {\rm and},\qquad
{\cal
H}=\sqrt{q}\,\left[R^{(3)}+q^{-1}(\half{\tilde{\pi}^2}-\tilde{\pi}^{ab}
\tilde{\pi}_{ab})\right]\approx 0 \ee
The first set of constraints are known as the vector constraint
and what they generate (its gauge orbit) are spatial
diffeomorphisms on $\Sigma$. The other constraint, the scalar
constraint (or super-Hamiltonian) generates ``time
reparametrizations". We start with 12 degrees of freedom, minus 4
constraints means that the constraint surface has 8 dimensions
(per point) minus the four gauge orbits generated by the
constraints giving the four phase space degrees of freedom, which
corresponds to the two polarizations of the gravitational field.

\subsection{Connection Dynamics}

In order to arrive at the connection formulation, we need first to enlarge
the phase space $\G_{\rm g}$ by considering not metrics $q_{ab}$ but the
 co-triads $e_a^i$ that define the metric by,
\be q_{ab}=e_a^i\,e_b^j\,\delta_{ij} \ee where $i,j=1,2,3$ are
internal labels for the frames. These represent 9 variables
instead of the 6 defining the metric $q_{ab}$, so we have
introduced more variables, but  at the same time a new symmetry in
the theory, namely the $SO(3)$ rotations in the triads. Recall
that a triad $e_a^i$ and a rotated triad $e^{\prime
i}_a(x)={U^i}_j(x)\,e^j_a(x)$ define the same metric $q_{ab}(x)$,
with ${U^i}_j(x) \in SO(3)$ a local rotation. In order to account
for the extra symmetry, there will be extra constraints (first
class) that will get rid of the extra degrees of freedom
introduced. Let us now introduce the densitized triad as follows:
\be
\tilde{E}^a_i=\half\,\epsilon_{ijk}\tilde{\eta}^{abc}\,e^j_b\,e^k_c
\ee
where ${\eta}^{abc}$ is the naturally defined levi-civita density
one antisymmetric object. Note that
$\tE^a_i\tE^b_j\delta^{ij}=q\,q^{ab}$.

Let us now consider the canonical variables. It turns out that the
canonical momenta to the densitized triad $\tE^a_i$ is closely
related to the extrinsic curvature of the metric, \be
K^i_a=\frac{1}{\sqrt{{{\rm
det}(\tE)}}}\;\delta^{ij}\tE^b_j\,K_{ab} \ee For details see
\cite{Perez:2004hj}. Once one has enlarged the phase space from
the pairs $(q_{ab},\tilde{\pi}^{ab})$ to $(\tE^a_i,K_b^j)$, the
next step is to perform the canonical transformation to go to the
$Ashtekar-Barbero$ variables. First we need to introduce the so
called {\it spin connection} $\G^i_a$, the one defined by the
derivative operator that annihilates the triad $e_a^i$ (in
complete analogy to the Christoffel symbol that defined the
covariant derivative $D_a$ killing the metric). It can be inverted
from the form,
\be
\partial_{[a}e^i_{b]} +{\epsilon^i}_{jk}\,\G^j_a\, e^k_b=0
\ee This can be seen as an extension of the covariant derivative
to objects with mixed indices. The key to the definition of the
new variables is to combine these two objects, namely the spin
connection $\G$ with the object $K^i_a$ (a tensorial object), to
produce a new connection
\be {}^{\gamma}\!A_a^i:=\G^i_a+\gamma\,K^i_a
\ee
This is the {\it Ashtekar-Barbero Connection}. Similarly, the
other conjugate variable will be the rescaled triad,
\be
{}^{\gamma}\!\tE^a_i=\tE^a_i/\gamma
\ee
Now, the pair $({}^{\gamma}\!A_a^i, {}^{\gamma}\!\tE^a_i)$ will
coordinatize the new phase space $\G_\gamma$. We have emphasized
the parameter $\g$ since this labels a one parameter family of
different classically equivalent theories, one for each value of
$\gamma$. The real and positive parameter $\g$ is known as the
Barbero-Immirzi parameter \cite{barbero,Immirzi}. In terms of
these new variables, the canonical Poisson brackets are given by,
\be \{ {}^{\gamma}\!A_a^i(x),
{}^{\gamma}\!\tE^b_j(y)\}=\kappa\,\delta^b_a\,\delta^i_j\,\delta^3(x,y)\,
. \ee and, \be \{ {}^{\gamma}\!A_a^i(x),
{}^{\gamma}\!A_b^j(y)\}=\{{}^{\gamma}\!\tE^a_i(x),
{}^{\gamma}\!\tE^b_j(y)\}=0
\ee
The new constraint that arises because of the
introduction of new degrees of freedom takes a very simple form,
\be
      G_i={\cal D}_a\,\tE^a_i\approx 0
\ee
that is, it has the structure of Gauss' law in Yang-Mills theory.
We have denoted
by ${\cal D}$ the covariant defined by the connection
${}^{\gamma}\!A_a^i$, such that ${\cal
D}_a\tE^a_i=\partial_a\tE^a_i+{\epsilon_{ij}}^{k}\;{}^{\gamma}
\!A_a^j\tE^a_k$. The vector and scalar constraints now take the
form,
\be
   V_a=F^i_{ab}\,\tE^b_i-(1+\g^2)K^i_a\,G_i \approx 0
\ee
where
$F^i_{ab}=\partial_{a}{}^{\gamma}\!A_b^i-\partial_{b}{}^{\gamma}\!A_a^i+
{\epsilon^i}_{jk}\;{}^{\gamma}\!A_a^j\,{}^{\gamma}\!A_b^k$ is the
curvature of the connection ${}^{\gamma}\!A_b^j$. The other
constraint is, \be \S=\frac{\tE^a_i\tE^b_j}{\sqrt{{\rm
det}(\tE)}}\,\left[{\epsilon^{ij}}_k\,F^k_{ab}-2(1+\gamma^2)
\,K^i_{[a}K^j_{b]}\right]\approx 0 \ee
The next step is to consider the right choice of variables, now
seen as functions of the phase space $\G_\g$ that are preferred
for the non-perturbative quantization we are seeking. As we shall
see, the guiding principle will be that the functions (defined by
an appropriate choice of smearing functions) will be those that
can be defined without the need of a background structure, i.e. a
metric on $\Sigma$.

\subsection{Holonomies and Fluxes}

\noindent  Since the theory possesses these constraints, the
strategy to be followed is to quantize first and then to impose
the set of constraints as operators on a Hilbert space. This is
known as the Kinematical Hilbert Space ${\cal H}_{\rm kin}$. One
of the main achievements of LQG is that this space has been
rigourously defined.

Let us start by considering the connection $A_a^i$. The most
natural object one can construct from a connection is a holonomy
$h_\a(A)$ along a loop $\a$. This is an element of the gauge group
$G=SU(2)$ and is denoted by,
\be h_\a(A)={\cal P}\,\exp\,\left( \oint_\a A_a\,\d s^a \right)
\ee
The path-order exponential of the connection. Note that for
notational simplicity we have omitted the `lie-algebra indices'.
 From the holonomy, it
is immediate to construct a gauge invariant function by taking the
trace arriving then at the Wilson loop $T[\a] := \frac{1}{2}\,
{\rm Tr}\,\, {\cal P}\,\exp\,(\oint_\a A_a\, \d s^a)$.

In recent years the emphasis has shifted from loops to consider
instead closed graphs $\Upsilon$, that consist of $N$ edges  $e_I$
($I=1,2,\ldots, N$), and $M$ vertices $v_\mu$, with the
restriction that there are no edges with `loose ends'.  Given  a
graph $\Upsilon$, one can consider the parallel transport along
the edges $e_I$, the end result is an element of the gauge group
$g_I=h(e_I)\in G$ for each such edge. One can then think of the
connection $A^i_a$ as a map from graphs to $N$-copies of the gauge
group: $A^i_a: \Upsilon\rightarrow G^N$. Furthermore, one can
think of ${\cal A}_{\Upsilon}$ as the configuration space  for the
graph $\Upsilon$, that is homeomorphic to $G^N$.

 What we are doing
at the moment is to construct relevant configuration functions. In
particular, what we need is to consider generalizations of the
Wilson loops $T[\a]$ defined previously, making use of the graphs
and the space ${\cal A}_{\Upsilon}$. Every graph $\Upsilon$ can be
decomposed into independent loops $\a_i$ and the corresponding
Wilson loops $T[\a_i]$ are a particular example of functions
defined over ${\cal A}_{\Upsilon}$. What we shall consider as a
generalization of the Wilson loop are {\it all} possible functions
defined over ${\cal A}_{\Upsilon}$. Thus, a function
$c:G^N\rightarrow \mathbb{C}$ defines a {\it cylindrical} function
$C_\Upsilon$ of the connection $A$ as,
  \be
C_\Upsilon:= c(h(e_1),h(e_2),\ldots,h(e_N))
  \ee
By considering all possible functions $c$ and all possible
embedded graphs $\Upsilon$, we generate the algebra of functions
known as Cyl (it is closely related to the holonomy algebra, and
it can be converted into a $C^*$-algebra $\overline{\rm Cyl}$, by
suitable completion).

 Let us now discuss why this choice
of configuration functions is compatible with the basic guiding
principles for the quantization we are building up, namely
diffeomorphism invariance and background independence. Background
independence is clear since there is no need for a background
metric to define the holonomies. Diffeomorphism invariance is a
bit more subtle. Clearly, when one applies a diffeomorphism
$\phi:\Sigma\rightarrow \Sigma$, the holonomies transform in a
covariant way
\be
  \phi_* \cdot h(e_I)=h(\phi^{-1}\cdot e_I)\, ,
\ee
that is, the diffeomorphism acts by moving the edge (or loop). How
can we then end up with a diffeo-invariant quantum theory? The
strategy in LQG is to look for a {\it diffeomorphism invariant}
representation of the diffeo-covariant configuration functions. As
we shall see later, this has indeed been possible and in a sense
represents the present `success' of the approach.

Let us now consider the functions depending of the momenta that
will be fundamental in the (loop) quantization. The basic idea is
again to look for functions that are defined in a background
independent way, that are natural from the view point of the
geometric character of the object (1-form, 2-form, etc), and that
transform covariantly with respect to the gauge invariances of the
theory. Just as the the connection $A_a^i$ can be identified with
a one form that could be integrated along a one-dimensional
object, one would like analyze the geometric character of the
densitised triad $\tE$ in order to naturally define a smeared
object. Recall that the momentum is a density-one vector field on
$\Sigma$, $\tE^a_i$ with values in the dual of the lie-algebra
$su(2)$. In terms of its tensorial character, it is naturally dual
to a (lie-algebra valued) two form,
\be
 E_{ab\;i}:=\half\, \teta_{abc}\,\tE^c_i
\ee
where $\teta_{abc}$ is the naturally defined Levi-Civita symbol.
It is now obvious that the momenta is crying to be integrated over
a two-surface $S$. It is now easy to define the objects
\be
 E[S,f]:=\int_S \;E_{ab\;i}\,f^i\,\d S^{ab}\, ,
\ee
where $f^i$ is a lie-algebra valued smearing function on $S$. This
`Electric flux' variable does not need a background metric to be
defined, and it transforms again covariantly as was the case of
the holonomies. The algebra generated by holonomies and flux
variables is known as the {\it Holonomy-Flux} algebra ${\cal HF}$.

Perhaps the main reason why this Holonomy-Flux  algebra
${\cal HF}$ is interesting, is the way in which the basic
generators interact, when considering the classical (Poisson)
lie-bracket. First, given that the configuration functions depend
only on the connection and the connections Poisson-commute, one
expects that $\{T[\a],T[\beta]\}=0$ for any loops $\a$ and
$\beta$. The most interesting poisson bracket one is interested in
is the one between a configuration and a momenta variable,
\be \{T[\a],E[S,f]\}= \kappa\,\sum_\mu
f^i(v)\,\iota(\a,S|_{v})\,{\rm Tr} \,(\tau_i\, h(\a))
\label{poissonB}
 \ee
where the sum is over the vertices $v$ and
$\iota(\a,S|_{v})=\pm 1$ is something like the
intersection number between the loops $\a$ and the surface $S$ at
point $v$. The sum is over all intersection of the loop $\a$
and the surface $S$. The most important property of the Poisson
Bracket is that it is completely topological. This has to be so if
we want to have a fully background independent classical algebra
for the quantization.

A remark is in order. The value of the constant $\iota|_{v}$
depends not only on the relative orientation of the tangent vector
of the loop $\a$ with respect to the orientation of $\Sigma$ and
$S$, but also on a further decomposition of the loop into edges,
and whether they are `incoming' or `outgoing' to the vertex $v$.
The end result is that is we have, for simple intersections, that
the number $\iota|_{v}$ becomes insensitive to the `orientation'.
This is different to the $U(1)$ case where the final result {\it
is} the intersection number. For details see \cite{ACZ}.

Let us now consider the slightly more involved case of a
cylindrical function $C_\Upsilon$ that is defined over a graph
$\Upsilon$ with edges $e_I$, intersecting the surface $S$ at
points $p$. We have then,
\be \{ C_\g, \, E[S,f] \} = \frac{\kappa}{2} \sum_{p}\,
\sum_{I_p}\, \iota({I_p})\, f^i(p)\, X^i_{I_p}\cdot c
\label{poissonbb}
 \ee
where the sum is over the vertices $p$ of the graph that lie on
the surface $S$, $I_p$ are the edges starting or finishing in $p$
and where $X^i_{I_P}\cdot c$ is the result of the action of the
$i$-th left (resp. right) invariant vector field on the $I_p$-th
copy of the group if the $I_p$-th edge is pointing away from
(resp. towards) the surface $S$.  Note the structure of the right
hand side. The result is non-zero only if the graph $\Upsilon$ used in
the definition of the configuration variable $C_\Upsilon$ intersects the
surface $S$ used to smear the triad. If the two intersect, the
contributions arise from the action of right/left invariant vector
fields on the arguments of $c$ associated with the edges at the
intersection.

Finally, the next bracket we should consider is between two
momentum functions, namely $\{E[S,f],E[S^\prime,g]\}$. Just as in
the case of holonomies, these functions depend only on one of the
canonical variables, namely the triad $\tE$. One should then
expect that their Poisson bracket vanishes. Surprisingly, this is
{\it not} the case and one has to appropriately define the correct
algebraic structure\footnote{The end result is that one should not
regard $E[S,f]$ as phase space functions subject to the ordinary
Poisson bracket relations, but rather should be viewed as arising
from vector fields $X^\alpha$ on ${\cal A}$. The non-trivial
bracket is then due to the non-commutative nature of the
corresponding vector fields. This was shown in \cite{ACZ} where
details can be found}.

We have arrived then to the basic variables that will be used in the
quantization in order to arrive at LQG. They are given by,
\be
h(e_I)\qquad {\rm Configuration \; function}
\ee
and
\be
  E[S,f]\qquad {\rm Momentum \; function},
\ee subject to the basic Poisson bracket relations given by
Eqs.~(\ref{poissonB}) and (\ref{poissonbb}). In the next section
we shall take the Holonomy-Flux algebra ${\cal HF}$ as the
starting point for the quantization.

\section{Loop Quantum Geometry}
\label{sec:3}

Let us now consider the particular representation that defines
LQG. As we have discussed before, the basic observables are
represented as operators acting on wave functions
$\Psi_\Upsilon(\overline{A})\in{\cal H}_{\rm kin}$ as follows:
\be \hat{h}(e_I)\cdot
\Psi(\overline{A})=\left(h(e_I)\;\Psi\right)(\bar{A})
 \ee
and
\be\label{11}
\hat{E}[S,f]\cdot\Psi_\Upsilon(\overline{A})=i\hbar\, \{
\Psi_\Upsilon, \, {}^2\!E[S,f] \} = i\,8\pi\,\frac{\ell^2_{\rm
P}}{2} \sum_{p}\, \sum_{I_p}\, \kappa({I_p})\, f^i(p)\,
X^i_{I_p}\cdot \psi \ee
where $\ell^2_{\rm P}=G\,\hbar$, the Planck area is giving us the
scale of the theory (recall that the Immirzi parameter $\g$ does
not appear in the basic Poisson bracket, and should therefore not
play any role in the quantum representation). Here we have assumed
that a cylindrical function $\Psi_\Upsilon$ on a graph $\Upsilon$
is an element of the Kinematical Hilbert space (which we haven't
defined yet!). This implies one of the most important assumption
in the loop quantization prescription, namely, that objects such
as holonomies and Wilson loops that are smeared in one dimension
are well defined operators on the quantum theory\footnote{this has
to be contrasted with the ordinary Fock representation where such
objects do {\it not} give raise to well defined operators on Fock
space. This implies that the loop quantum theory is qualitatively
different from the standard quantization of gauge fields.}.

The basic idea for the construction of both the Hilbert space
${\cal H}_{\rm kin}$ (with its measure) and the quantum
configuration space $\bar{\cal A}$, is to consider {\it all}
possible graphs on $\Sigma$. For any given graph $\Upsilon$, we
have a configuration space ${\cal A}_\Upsilon=(SU(2))^N$, which is
$n$-copies of the (compact) gauge group $SU(2)$. Now, it turns out
that there is a preferred (normalized) measure on any compact
semi-simple Lie group that is left and right invariant. It is
known as the Haar measure $\mu_{\rm H}$ on the group. We can thus
endow ${\cal A}_\Upsilon$ with a measure $\mu_\Upsilon$ that is
defined by using the Haar measure on all copies of the group.
Given this measure on ${\cal A}_\Upsilon$, we can consider square
integrable functions thereon and with them the graph-$\Upsilon$
Hilbert space ${\cal H}_\Upsilon$, which is of the form:
\be
 {\cal H}_\Upsilon=L^2({\cal A}_\Upsilon,\d\mu_\Upsilon)
\ee
If we were working with a unique, fixed graph $\Upsilon_0$ (which
we are not), we would be in the case of a lattice gauge theory on
an irregular lattice. The main difference between that situation
and LQG is that, in the latter case, one is considering all graphs
on $\Sigma$, and one has a family of configurations spaces $\{
{\cal A}_\Upsilon \,/ \Upsilon \,{\rm a\,graph\,in\,} \Sigma\}$,
and a family of Hilbert spaces $\{ {\cal H}_\Upsilon \,/ \Upsilon
\,{\rm a\,graph\,in\,} \Sigma\}$.

The quantum configuration space $\overline{\cal A}$ is the
configuration space for the ``largest graph"; and similarly, the
kinematical Hilbert space ${\cal H}_{\rm kin}$ is the largest
space containing all Hilbert spaces in $\{ {\cal H}_\Upsilon \,/
\Upsilon \,{\rm a\,graph\,in\,} \Sigma\}$.  The
Ashtekar-Lewandowski  measure $\mu_{\rm AL}$ on ${\cal H}_{\rm
kin}$ is then the measure whose projection to any ${\cal
A}_\Upsilon$ yields the corresponding Haar measure $\mu_\Upsilon$.
The resulting Hilbert space can thus be written as
$$
{\cal H}_{\rm kin}=L^2(\overline{\cal A},\d\mu_{\rm AL})
$$
The cylindrical functions $\Psi_\Upsilon\in$ Cyl belong to the
Hilbert space of the theory.

Let us then recall what is the structure of simple states in the
theory. The vacuum or `ground state' $|0\rangle$ is given by the
unit function. One can then create excitations by acting via
multiplication with holonomies or Wilson loops. The resulting
state $|\a\rangle=\hat{T}[\a]\cdot|0\rangle$ is an excitation of
the geometry but only along the one dimensional loop $\a$. Since
the excitations are one dimensional, the geometry is sometimes
said to be {\it polymer like}. In order to obtain a geometry that
resembles a three dimensional continuum one needs a huge number of
edges ($10^{68}$) and vertices.

\subsection{A choice of basis: Spin Networks}

The purpose of this part is to provide a useful decomposition of
the Hilbert space ${\cal H}_\Upsilon$, for all graphs. From our
previous discussion we know that the Hilbert space ${\cal
H}_\Upsilon$ is the Cauchy completion of the space of cylinder
functions on $\Upsilon$, ${\rm Cyl}_\Upsilon$ with respect to the
norm induced by the Haar measure on the graph configuration space
${\cal A}_\Upsilon = (\SU(2))^N$. Thus, what we are looking for is
a convenient basis for functions $F_\Upsilon(A)$ of the form
\be F_\Upsilon(A):= f(h(e_1),h(e_2),\ldots,h(e_N)) \ee

Let us for a moment consider just one edge, say $e_i$. What we
need to do is to be able to decompose any function $F$ on $G$ (in
this case we only have one copy of the group), in a suitable
basis.

In the case of the group $G=SU(2)$, there is a decomposition of a
function $f(g)$ of the group ($g\in G$). It reads,
\be
 f(g)=\sum_j\sqrt{j(j+1)}\,f^{m m'}_j\stackrel{j}{\Pi}_{m m'}\;(g)
\ee where, \be f^{m m'}_j=\sqrt{j(j+1)}\int_G\d\mu_H
\stackrel{j}{\Pi}_{m m'}(g^{-1})\,f(g)
\ee
and is the equivalent of the Fourier component. The functions
$\stackrel{j}{\Pi}_{m m'}\,(g)$ play the role of the Fourier
basis. In this case these are unitary representation of the group,
and the label $j$ labels the irreducible representations. In the
$SU(2)$ case with the interpretation of spin, these represent the
spin-$j$ representations of the group. In our case, we will
continue to use that terminology (spin) even when the
interpretation is somewhat different.

Given a cylindrical function
$\Psi_\Upsilon[A]=\psi(h(e_1),h(e_2),\ldots,h(e_N))$, we can then
write an expansion for it as,
\ba
\Psi_\Upsilon[A] &=& \psi(h(e_1),h(e_2),\ldots,h(e_N))\nonumber\\
& = & \sum_{j_1\cdots j_N}f^{m_1\cdots m_N,n_1\cdots
n_N}_{j_1\cdots j_N}
\;\phi^{j_1}_{m_1n_1}(h(e_1))\cdots\phi^{j_N}_{m_Nn_N}(h(e_N)),
\ea where $\phi^{j}_{mn}(g)=\sqrt{j(j+1)}\,\stackrel{j}{\Pi}_{m
n}(g)$ is the normalized function satisfying
$$
\int_G \d\mu_{\rm H}\,\overline{\phi^{j}_{mn}(g)}\;\phi^{j'}_{m'
n'}(g)= \delta_{j,j'} \delta_{m,m'}\delta_{n,n'}\, .
$$
The expansion coefficients can be obtained by projecting the state
$|\Psi_\Upsilon\rangle$,
\be f^{m_1\cdots m_N,n_1\cdots n_N}_{j_1\cdots j_N} =\langle
\,\phi^{j_1}_{m_1n_1}\cdots\phi^{j_N}_{m_Nn_N}\;|\;\Psi_\Upsilon
\rangle \ee
This implies that the products of components of irreducible
representations $\prod^{N}_{i=1}\phi^{j_i}_{m_i n_i}[h(e_i)]$
associated with the $N$ edges $e_I\in\Upsilon$, for all values of
spins $j$ and for $-j\leq m,n\leq j$ and for any graph $\Upsilon$,
is a complete orthonormal basis for ${\cal H}_{\rm kin}$. We can
the write, \be {\cal H}_\Upsilon=\otimes_j\;{\cal H}_{\Upsilon,j}
\ee where the Hilbert space ${\cal H}_{\a,j}$ for a single loop
$\a$, and a label $j$ is the familiar $(2j+1)$ dimensional Hilbert
space of a particle of `spin $j$'. For a complete treatment see
\cite{Perez:2004hj}.

In the case of geometry with group $SU(2)$, the graphs with
labelling $j_I={\bf j}$ are known as spin networks. As the reader
might have noticed, in the geometry case there are more labels
than the spins for the edges. Normally these are associated to
vertices and are known as intertwiners. This means that the
Hilbert spaces ${\cal H}_{\Upsilon,{\bf j}}$ is finite
dimensional. Its dimension being a measure of the extra freedom
contained in the intertwiners. One could then introduce further
labelling $\bf l$ for the graph, so we can decompose the Hilbert
space as
\be {\cal H}_\Upsilon=\otimes_j\;{\cal H}_{\Upsilon,{\bf
j}}=\otimes_{{\bf j},{\bf l}}\;{\cal H}_{\Upsilon,{\bf j},{\bf l}}
\ee
where now the spaces ${\cal H}_{\Upsilon,{\bf j},{\bf l}}$ are
one-dimensional. For more details see \cite{ac:playa},
\cite{Perez:2004hj} and \cite{AL:review}. With this convenient
basis it is simple to consider geometrical operator. The most
important one in the study of black holes is given by the flux and
the area operators that we consider next.

\subsection{Flux and Area operators}

The operators $\hat{E}[S,f]$ corresponding to the electric flux
observables, are in a sense the basic building blocks for
constructing the quantum geometry. We have seen in Sec.\ref{sec:3}
the action of this operators on cylindrical functions,
\be
 \hat{E}[S,f]\cdot\Psi_\Upsilon(\overline{A})=-i \,\hbar\,
\{ \Psi_\Upsilon, \, {}^2\!E[S,f] \} = -i\,8\pi\,\frac{\ell^2_{\rm
P}}{2} \sum_{p}\, \sum_{I_p}\, \kappa({I_p})\, f^i(p)\,
X^i_{I_p}\cdot \psi\label{flux1} \ee
Here the first sum is over the intersections of the surface $S$
with the graph $\Upsilon$, and the second sum is over all possible
edges $I_p$ that have the vertex $v_p$ (in the intersection of $S$
and the graph) as initial of final point. In the simplest case of
a loop $\a$, there are only simple intersections (meaning that
there are two edges for each vertex), and in the simplest case of
only one intersection between $S$ and $\a$ we have one term in the
first sum and two terms in the second (due to the fact that the
loop $\a$ is seen as having a vertex at the intersection point).
In this simplified case we have
\be
 \hat{E}[S,f]\cdot\Psi_\a(\overline{A})=\,
-i\,8\pi\,\ell^2_{\rm P} \,   f^i(p)\, X^i_{I_p}\cdot \psi
 \ee
Note that the action of the operator is to `project' the angular
momentum in the direction given by $f^i$ (in the internal space
associated with the Lie algebra). As we shall see, this operator
is in a sense fundamental the fundamental entity for constructing
(gauge invariant) geometrical operators. For this, let us rewrite
the action of the flux operator (\ref{flux1}), dividing the edges
that are above the surface $S$, as `up' edges, and those that lie
under the surface as `down' edges. \be
\hat{E}[S,f]\cdot=8\pi\frac{\ell^2_P}{2}\sum_p
f^i(p)\,(\hat{J}^{p}_{i(u)}- \hat{J}^{p}_{i(d)})\cdot \ee where
the sum is over the vertices at the intersection of the graph and
the surface, and where the `up' operator
$\hat{J}^{p}_{i(u)}=\hat{J}^{p,e_1}_{i}+
\hat{J}^{p,e_2}_{i}+\cdots+\hat{J}^{p,e_u}_{i}$ is the sum over
all the up edges and the down operator  $\hat{J}^{p}_{i(d)}$ is
the corresponding one for the down edges.

The second simplest operator that can be constructed representing
geometrical quantities of interest is the {\it area operator},
associated to surfaces $S$. The reason behind this is again the
fact that the densitized triad is dual to a two form that is
naturally integrated along  a surface. The difference between the
classical expression for the area and the flux variable is the
fact that the area is a gauge invariant quantity. Let us first
recall what the classical expression for the area function is, and
then we will outline the regularization procedure to arrive at a
well defined operator on the Hilbert space. The area $A[S]$ of a
surface $S$ is given by $A[S]=\int_S\d^2 x\,\sqrt{h}$, where $h$
is the determinant of the induced metric $h_{ab}$ on $S$. When the
surface $S$ can be parametrized by setting, say, $x^3=0$, then the
expression for the area in terms of the densitized triad takes a
simple form:
 \be
A[S]=\gamma\int_S\d^2
x\,\sqrt{\tilde{E}^3_i\,\tilde{E}^3_j\,k^{ij}}
 \ee
where $k^{ij}=\delta^{ij}$ is the Killing-Cartan metric on the Lie
algebra, and $\gamma$ is the Barbero-Immirzi parameter (recall
that the canonical conjugate to $A$ is
${}^\gamma\!\tilde{E}^a_i=\tilde{E}^a_i/\gamma$). Note that the
functions is again smeared in two dimensions and that the quantity
inside the square root is very much a square of the (local) flux.
One expects from the experience with the flux operator, that the
resulting operator will be a sum over the intersecting points $p$,
so one should focus the attention to the vertex operator
$$\Delta_{S,\Upsilon,p}=-\left[(\hat{J}^{p}_{i(u)}-
\hat{J}^{p}_{i(d)})(\hat{J}^{p}_{j(u)}-
\hat{J}^{p}_{j(d)})\right]k^{ij}$$ with this, the area operator
takes the form,
\be \hat{A}[S]=8\pi\,\gamma\,\ell^2_{\rm P}\sum_p
\widehat{\sqrt{\Delta_{S,\Upsilon,p}}} \ee We can now combine both
the form of the vertex operator with Gauss' law
$(\hat{J}^{p}_{i(u)}+ \hat{J}^{p}_{i(d)})\cdot \Psi=0$ to arrive
at,
\be |(\hat{J}^{p}_{i(u)} - \hat{J}^{p}_{i(d)})|^2 =
|2(\hat{J}^{p}_{i(u)})|^2 \ee
where we are assuming that there are no tangential edges. The
operator $\hat{J}^{p}_{i(u)}$ is an angular momentum operator, and
therefore its square has eigenvalues equal to $j^u(j^u+1)$ where
$j^u$ is the label for the total `up' angular momentum. We can
then write the form of the operator \be \hat{A}[S]\cdot{\cal
N}(\Upsilon,\vec{j})=8\pi\,\gamma\,\ell^2_{\rm P}
 \sum_{v\in \,V}
 \;\sqrt{|\hat{J}^{p}_{i(u)}|^2}\cdot
{\cal N}(\Upsilon,\vec{j}) \ee
With these conventions, in the case of  simple intersections
between the graph $\Upsilon$ and the surface $S$, the area
operator takes the well known form:
\[
\hat{A}[S]\cdot{\cal
N}(\Upsilon,\vec{j})=8\pi\,\gamma\,\ell^2_{\rm P} \sum_{v\in \,V}
 \;\sqrt{j_v(j_v+1)}\cdot
{\cal N}(\Upsilon,\vec{j})
\]
when acting on a {\it spin network} ${\cal N}(\gamma,\vec{j})$
defined over $\Upsilon$ and with labels $\vec{j}$ on the edges (we
have not used a label for the intertwiners). As we shall se when
we consider the quantum theory of isolated horizons, the two
operators considered here will play an important role not only in
the geometry of the horizon but in the entropy counting.

\section{Quantum Isolated Horizons}
\label{sec:4}

\noindent
Let us focus on the sector of the theory consisting of
space-times which admit a type I isolated horizon $\Delta$ with a
fixed area $a_o$ as the internal boundary. Then $\Sigma$ is
asymptotically flat and has an internal boundary $S$,
topologically a  2-sphere, the intersection of $\Sigma$ with
$\Delta$. The type I isolated horizon boundary conditions require
that i) $\Delta$ be null, ii) Non-expanding, iii) The field
equations be satisfied there and iv) the intrinsic geometry on
$\Delta$ be left invariant by the null vecto $\ell^a$ generating
$\Delta$. For details see \cite{Askris04}.

Introduce on $S$ an internal, unit, radial vector field $r^i$
(i.e. any isomorphism from the unit 2-sphere in the Lie algebra of
$\SU(2)$ to $S$). Then it turns out that \emph{the intrinsic
geometry of $S$ is completely determined by the pull-back
$\ub{A}^ir_i =: W$ to $S$ of the (internal-radial component of
the) connection $A^i$ on $\Sigma$} \cite{Askris04}. Furthermore,
$W$ is in fact a spin-connection intrinsic to the 2-sphere $S$.
Finally, the fact that $S$ is (the intersection of $\Sigma$ with)
a type I isolated horizon is captured in a relation between the
two canonically conjugate fields:
\be \label{bc1} F:\=\,\, \d W\,\, \=  - \f{2\pi\g}{a_o}\,\,
\underline\Sigma^i\, r_i .\ee
where $\underline{\Sigma}^i$ is the pull-back to $S$ of the
2-forms $\Sigma_{ab}^i=\eta_{abc}E^{i\,c}$ on $\Sigma$.
(Throughout, $\=$ will stand for equality restricted to $\Delta$.)
Thus, because of the isolated horizon boundary conditions, fields
which would otherwise be independent are now related. In
particular, the pull-backs to $S$ of the canonically conjugate
fields $A^i,\, \Sigma^i$ are completely determined by the $U(1)$
connection $W$.

In absence of an internal boundary, the symplectic structure is
given just by a volume integral \cite{Askris04}. In presence of
the internal boundary under consideration, it now acquires a
surface term \cite{ABCK}:
\be \label{sym1} {\bf \Omega}(\delta_1, \delta_2) = \f{1}{8\pi
G}\left[ \int_M \, {\rm  Tr}\, (\delta_1 A \wedge \delta_2 \Sigma
- \delta_2 A \wedge \delta_1 \Sigma) \, +\, \f{a_o}{\g \pi}
\oint_S \, \delta_1 W \wedge \delta_2 W\,   \right]\, , \ee
where $\delta \equiv (\delta A, \delta \Sigma)$ denotes tangent
vectors to the phase space ${\bf \Gamma}$. Since $W$ is
essentially the only `free data' on the horizon, it is not
surprising that the surface term of the symplectic structure is
expressible entirely in terms of $W$. However, it is interesting
that the new surface term is precisely the symplectic structure of
a $\U(1)$-Chern Simons theory. \emph{The symplectic structures of
the Maxwell, Yang-Mills, scalar and dilatonic fields do not
acquire surface terms and, because of minimal coupling, do not
feature in the gravitational symplectic structure either.}
Conceptually, this is an important point: this, in essence, is the
reason why the black hole entropy depends just on the horizon area
and not, in addition, on the matter charges \cite{ABCK}.

One can systematically `quantize' this sector of the phase space
\cite{ABCK}. We can focus only on the gravitational field since
the matter fields do not play a significant role. One begins with
a Kinematic Hilbert space $\H = \H_{V}\otimes \H_{S}$ where $\H_V$
is the Hilbert space of states in the bulk as described before and
$\H_S$ is the Hilbert space of surface states. Expression
(\ref{sym1}) of the symplectic structure implies that $\H_S$
should be the Hilbert space of states of a Chern-Simons theory on
the punctured $S$, where the `level', or the coupling constant, is
given by:
\be \label{level1} k = \f{a_o}{4\pi\gamma\lp^2} \ee
A pre-quantization consistency requirement is that $k$ be an
integer \cite{ABCK}.

Our next task is to encode in the quantum theory the fact that
$\Delta$ is a type I horizon with area $a_o$. This is done by
imposing the horizon boundary condition  as an \emph{operator
equation}:
\be \label{qbc1} (1\otimes \hat{F})\, \Psi\, \=\, -
\left(\frac{2\pi\gamma}{a_o}\, (\hat{\underline{\Sigma}}\cdot
r)\otimes 1\right)\, \Psi \, , \ee
on admissible states $\Psi$ in $\H$. Now, a general solution to
(\ref{qbc1}) can be expanded out in a basis: $\Psi = \sum_n\,
\Psi_V^{(n)} \otimes \Psi_S^{(n)}$, where $\Psi_V^{(n)}$ is an
eigenvector of the `triad operator' $- ({2\pi\gamma}/{a_o})\,
(\hat{\underline{\Sigma}}\cdot r)(x)$ on $\H_V$ and $\Psi_S^{(n)}$
is an eigenvector of the `curvature operator' $\hat{F}(x)$ on
$\H_S$ \emph{with same eigenvalues}. Thus, the theory is
non-trivial only if a sufficiently large number of eigenvalues of
the two operators coincide. Since the two operators act on
entirely different Hilbert spaces and are introduced quite
independently of one another, this is a \emph{very} non-trivial
requirement.

Now, in the bulk Hilbert space $\H_V$, the eigenvalues of the
`triad operator' are given by \cite{ACZ}:
\be  \label{triadev1} -\, \left(\f{2\pi\gamma}{a_o}\right)\,\,
\left(8\pi \lp^2\,\, \sum_I\, m_I \delta^3(x, p_I)\,
\eta_{ab}\right)\, , \ee
where $m_I$ are half integers and $\eta_{ab}$ is the natural,
metric independent Levi-Civita density on $S$ and $p_I$ are points
on $S$ at which the polymer excitations of the bulk geometry in
the state $\Psi_V$ puncture $S$. A completely independent
calculation \cite{ABCK}, involving just the surface Hilbert space
$\H_S$, yields the following eigenvalues of $\hat{F}(x)$:
\be \label{Fev1} \f{2\pi}{k}\, \sum_I\, n_I\, \delta^3(x, p_I)\,
\equiv \,  2\pi\, \f{4\pi \gamma\lp^2}{a_o}\, \sum_I\, n_I
\,\delta^3(x, p_I)\, \ee
where $n_I$ are integers modulo $k$. Thus, with the identification
$-2m_I = n_I\, {\rm mod}\, k$, the two sets of eigenvalues match
exactly. There is a further requirement or constraint that the
numbers $m_I$ should satisfy, namely,
\be \sum_I m_I=0 \label{46}
\ee
This constraint is sometimes referred to as the \emph{projection
constraint}, given that the `total projection of the angular
momentum' is zero. Note that in the Chern-Simons theory the
eigenvalues of $F(x)$ are dictated by the `level' $k$ and the
isolated horizon boundary conditions tie it to the area parameter
$a_o$ just in the way required to obtain a coherent description of
the geometry of the quantum horizon.

In the classical theory, the parameter $a_{o}$ in the expression
of the surface term of the symplectic structure (\ref{sym1}) and
in the boundary condition (\ref{bc1}) is the horizon area. However
in the \emph{quantum theory}, $a_{o}$ has simply been a parameter
so far; we have not tied it to the \emph{physical area of the
horizon}. Therefore, in the entropy calculation, to capture the
intended physical situation, one constructs a suitable
`micro-canonical' ensemble. This leads to the last essential
technical step.

Let us begin by recalling that, in quantum geometry, the area
eigenvalues are given by,
$$ 8\pi \gamma\lp^2\, \sum_I \sqrt{j_I(j_I +1)}\, . $$
We can therefore construct a micro-canonical ensemble by
considering only that sub-space of the volume theory which, at the
horizon, satisfies:
\be \label{micro1} a_{o} -\delta  \le 8\pi\gamma \lp^2\, \sum_I\,
\sqrt{j_I(j_I+1)} \le a_{o} + \delta \ee
where $I$ ranges over the number of punctures, $j_I$ is the spin
label associated with the puncture $p_I$ \cite{ABCK}.%
\footnote{The appearance of the parameter $\delta$ is standard in
statistical mechanics. It has to be much smaller than the
macroscopic parameters of the system but larger than level
spacings in the spectrum of the operator under consideration.}

Quantum Einstein's equations can be imposed as follows. The
implementation of the Gauss and the diffeomorphism constraints is
the same as in \cite{ABCK}. The first says that the `total' state
in $\H$ be invariant under the $\SU(2)$ gauge rotations of triads
and, as in \cite{ABCK}, this condition is automatically met when
the state satisfies the quantum boundary condition (\ref{qbc1}).
The second constraint says that two states are physically the same
if they are related by a diffeomorphism. The detailed
implementation of this condition is rather subtle because an extra
structure is needed in the construction of the surface Hilbert
space and the effect of diffeomorphisms on this structure has to
be handled carefully \cite{ABCK}. However, the final result is
rather simple: For surface states, what matters is only the number
of punctures; their location is irrelevant. The last quantum
constraint is the Hamiltonian one. In the classical theory, the
constraint is differentiable on the phase space only if the lapse
goes to zero on the boundary. Therefore, this constraint restricts
only the volume states. However, there is an indirect restriction
on surface states which arises as follows. Consider a set $(p_I,
j_I)$ with $I= 1,2,\ldots N$ consisting of $N$ punctures $p_I$ and
half-integers $j_I$, real, satisfying (\ref{micro1}). We will
refer to this set as `surface data'. Suppose there exists a bulk
state satisfying the Hamiltonian constraint which is compatible
with this `surface data'. Then, we can find compatible surface
states such that the resulting states in the total Hilbert space
$\H$ lie in our ensemble. The space $S_{(p_I,j_I)}$ of these
surface states is determined entirely by the surface data. In our
state counting, we include the number ${\cal N}_{(p_I,j_I)}$ of
these surface states, subject however, to the projection
constraint
that is purely intrinsic to the horizon.%
\footnote{Note that there may be a large number --possibly
infinite-- of bulk states which are compatible with a given
`surface data' in this sense. This number does not matter because
the bulk states are `traced out' in calculating the entropy of the
horizon. What matters for the entropy calculation is only the
dimensionality of $S_{(p_I, j_I)}$.}

\section{Black Hole Entropy}
\label{sec:5}

\noindent
 In this section we shall deal with the issue of entropy
counting. We have started with a type I isolated horizon of area
$a_o$ (in vacuum this is the only multipole defining the horizon),
and we have quantized the theory and arrived to a Hilbert space as
described before. The question now is: How many microstates
correspond to the given macrostate, defined uniquely by $a_o$?

Let us now pose the condition that the states in $S_{(p_I, j_I)}$
should satisfy:

\begin{itemize}

\item They belong to the \emph{physical} Hilbert space on the
surface $\H_S$.

\item The condition (\ref{micro1}) is satisfied.

\item The quantum boundary condition (\ref{Fev1})  is satisfied.

\item The projection constraint (\ref{46}) is satisfied.

\end{itemize}

In terms of a concrete counting the problem is posed a follows: We
shall consider the lists $(p_I, j_I, m_I)$ corresponding to the
allowed punctures, spins of the piercing edges, and `projected
angular momentum' labels, respectively.

The task is then to count these states and find ${\cal
N}_{(p_I,j_I, m_I)}$. The entropy will be then,
\be
 S_{\rm BH}:= \ln ({\cal N}_{(p_I,j_I, m_I)})
\ee
This problem was systematically addressed in \cite{ABCK} in the
approximation of \emph{large} horizon area $a_o$. Unfortunately,
the number of such states was underestimated in \cite{ABCK}.%
\footnote{The under-counting was noticed in \cite{Dom:Lew}, and a
new counting was there proposed and carried out in \cite{meiss}.
However the choice of relevant states to be counted there is
slightly different from our case. Details of the comparison
between two methods will be reported elsewhere \cite{CDF-3}.} In
\cite{GM} the counting was completed and it was shown in detail
that, for large black holes (in Planck units), the entropy behaves
as:
$$S_{\rm BH}=\frac{A}{4}-\frac{1}{2}\ln{A},$$ provided the Barbero-Immirzi
parameter $\gamma$ is chosen to coincide with the value
$\gamma_0$, that has to satisfy \cite{GM}: \be
 1=\sum_I
(2j_I+1)\,\exp{\left(-2\pi\gamma_0\sqrt{j_I(j_I+1)}\right)}\, .
\ee The solution to this transcendental equation is approximately
$\gamma_0=0.27398\ldots$ (see \cite{GM,CDF-3} for details).

Here we shall perform the counting in a different regime, namely
for small black holes in the Planck regime \cite{CDF}. Thus we
shall perform no approximations as in the previous results. Thus
our results are complementary to the analytic treatments. On the
other hand our counting will be exact, since the computer
algorithm is designed to count all states allowed. Counting
configurations for large values of the area (or mass) is extremely
difficult for the simple reason that the number of states scales
exponentially. Thus, for the computing power at our disposal, we
have been able to compute states up to a value of area of about
$a_o=550\; l^2_P$ (recall that the minimum area gap for a spin
$1/2$ edge is about $a_{\rm min} \approx 6\,l^2_P$, so the number
of punctures on the horizon is below 100). At this point the
number of states exceeds $2.8\times 10^{58}$. In terms of Planck
masses, the largest value we have calculated is $M=3.3 \,M_P$.
When the projection constraint is introduced, the upper mass we
can calculate is much smaller, given the computational complexity
of implementing the condition. In this case, the maximum mass is
about $1.7\,M_P$.

It is important to describe briefly what the program for counting
does. What we are using is what it is known, within combinatorial
problems, as a brute force algorithm. This is, we are simply
asking the computer to perform all possible combinations of the
labels we need to consider, attending to the distinguishability
-indistinguishability criteria that are relevant
\cite{ABCK,CDF-3}, and to select (count) only those that satisfy
the conditions needed to be considered as permissible
combinations, i.e., the area condition and the spin projection
constraint. An algorithm of this kind has an important
disadvantage: it is obviously not the most optimized way of
counting and the running time increases rapidly as we go to little
higher areas. This is currently the main limitation of our
algorithm. But, on the other hand, this algorithm presents a very
important advantage, and this is the reason why we are using it:
its explicit counting guarantees us that, if the labels considered
are correct, the result must be the right one, as no additional
assumption or approximation is being made. It is also important to
have a clear understanding that the algorithm does not rely on any
particular analytical counting available. That is, the program
counts states as specified in the original formalism \cite{ABCK}.
The computer program has three inputs: i) the classical mass $M$
(or area $a_o=16\pi\,M^2$), ii) The value of $\gamma$ and iii) The
size of the interval $\delta$.

\begin{figure}
  \includegraphics[angle=270,scale=.40]{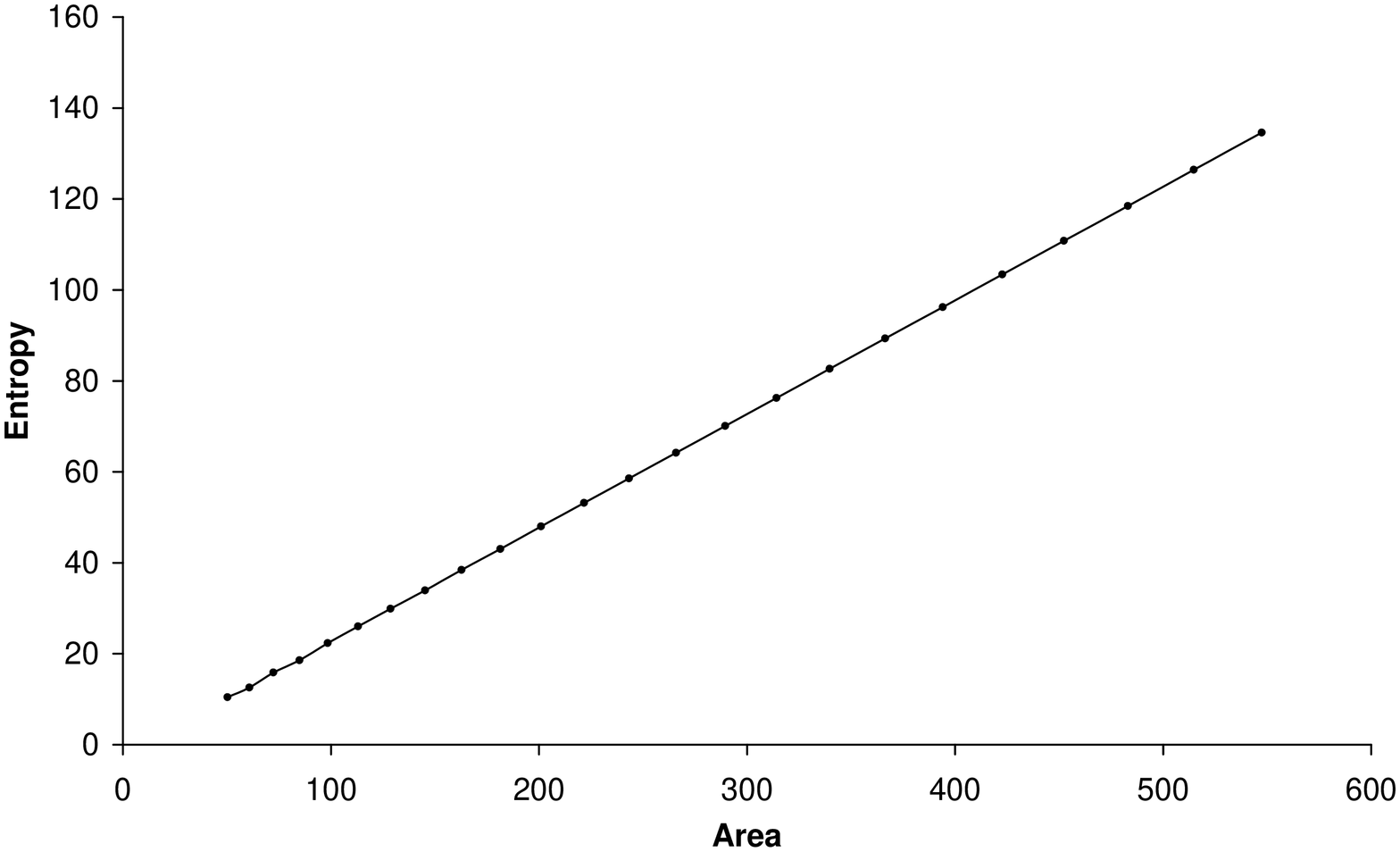}
\caption{\label{fig:1} The entropy as a function of area is shown,
where the projection constraint has not been imposed. The BI is
taken as $\gamma=0.274$.}
\end{figure}

Once these values are given, the algorithm computes the level of
the horizon Chern Simons theory $k=\left[a_o/4\pi\gamma\right]$
and the maximum number of punctures possible $n_{\rm
max}=\left[a_o/4\pi\gamma\sqrt{3}\right]$ (where $[\cdot]$,
indicates the principal integer value). At first sight we see that
the two conditions we have to impose to permissible combinations
act on different labels. The area condition acts over $j$'s and
the spin projection constraint over $m$'s. This allows us to first
perform combinations of $j$'s and select those satisfying the area
condition. After that, we can perform combinations of $m$'s only
for those combinations of $j$'s with the correct area, avoiding
some unnecessary work. We could also be allowed to perform the
counting without imposing the spin projection constraint, by
simply counting combinations of $j$'s and including a multiplicity
factor of $\prod_I (2 j_I + 1)$ for each one, accounting for all
the possible combinations of $m$'s compatible with each
combination of $j$'s. This would reduce considerably the running
time of the program, as no counting over $m$'s has to be done, and
will allow us to separate the effects of the spin projection
constraint (that, as we will see, is the responsible of a
logarithmic correction). It is very important to notice at this
point that this separation of the counting is completely
compatible with the distinguishability criteria.

The next step of the algorithm  is to consider, in increasing
order, all the possible number of punctures (from 1 to $n_{\rm
max}$) and in each case it considers all possible values of $j_I$.
Given a configuration $(j_1,j_2,\ldots,j_n)$ ($n\leq n_{\rm
max}$), we ask whether the quantum area eigenvalue $A=\sum_I
8\pi\gamma\sqrt{{j_I}\left({j_I}+1\right)}$ lies within
$[a_o-\delta ,a_o+\delta ]$. If it is not, then we go to the next
configuration. If the answer is positive, then the labels $m$'s
are considered as described before. That is, for each of them it
is checked whether $\sum m_I=0$ is satisfied.

\begin{figure}
  \includegraphics[angle=270,scale=.40]{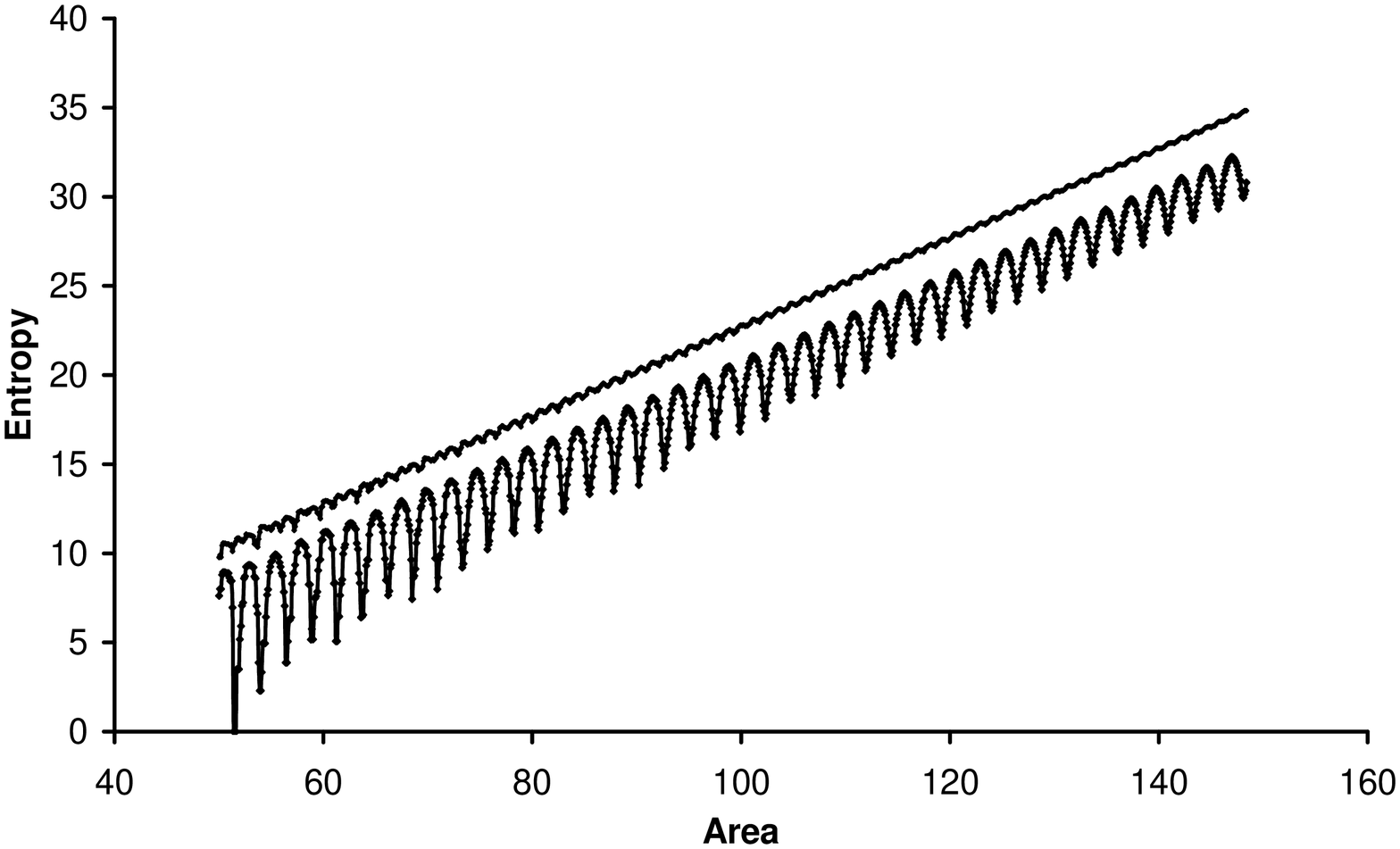}
\caption{\label{fig:2} Entropy {\it vs} Area with and without the
projection constraint, with $\delta =0.5$.}
\end{figure}

In Figure~\ref{fig:1}, we have plotted the entropy, as
$S=\ln(\#\,{\rm states})$ {\it vs} the area $a_o$, where we have
counted all possible states without imposing the $\sum m_I$
constraint, and have chosen a $\delta=0.5$. As it can be seen, the
relation is amazingly linear, even for such small values of the
area. When we fix the BI parameter to be $\gamma=\gamma_0=0.274$,
and approximate the curve by a linear function, we  find that the
best fit is for a slope equal to $0.2502$.

\begin{figure}
  \includegraphics[angle=270,scale=.40]{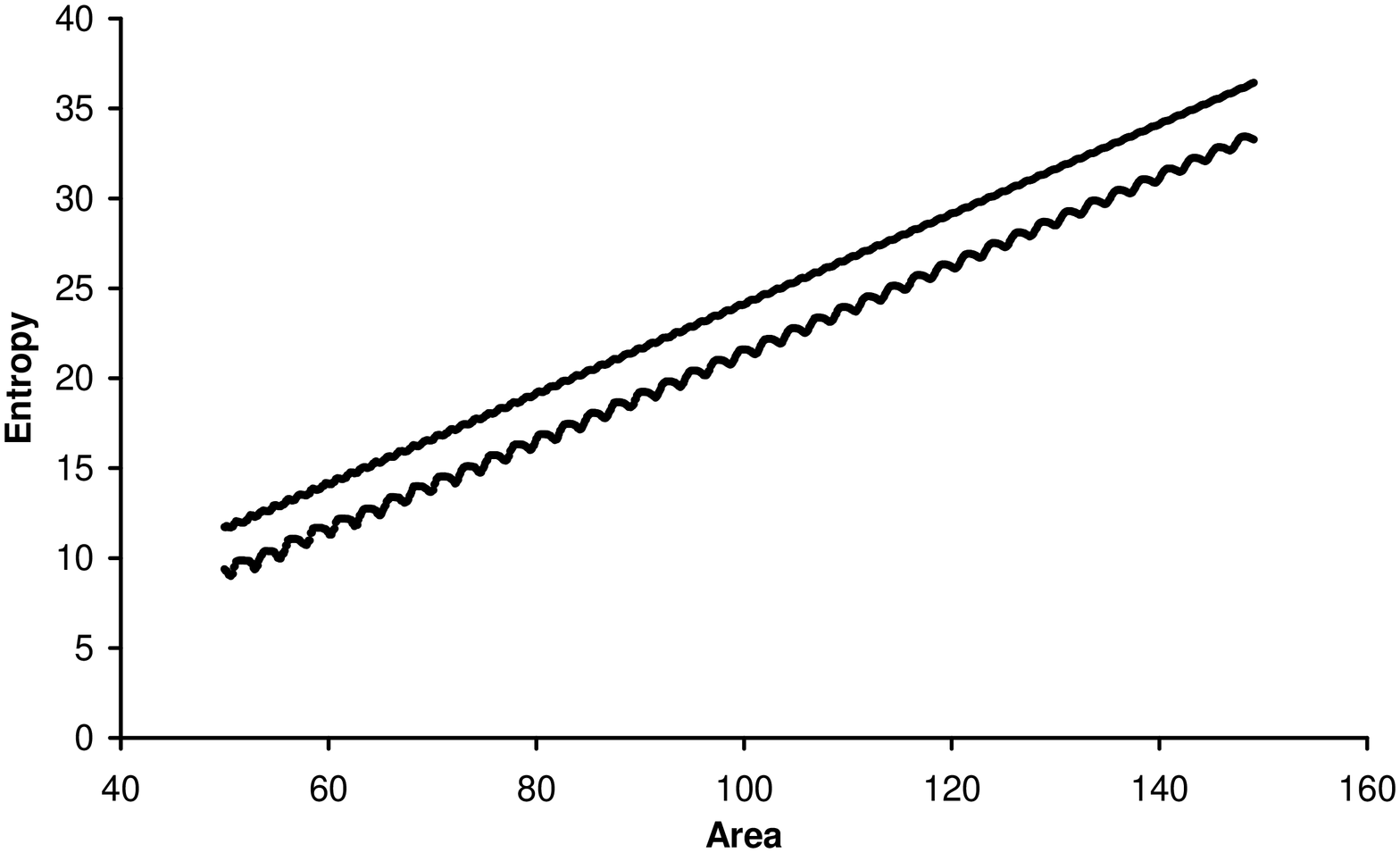}
\caption{\label{fig:3} Entropy {\it vs} Area with and without the
projection constraint, with $\delta =2$.}
\end{figure}
\begin{figure}
  \includegraphics[angle=270,scale=.39]{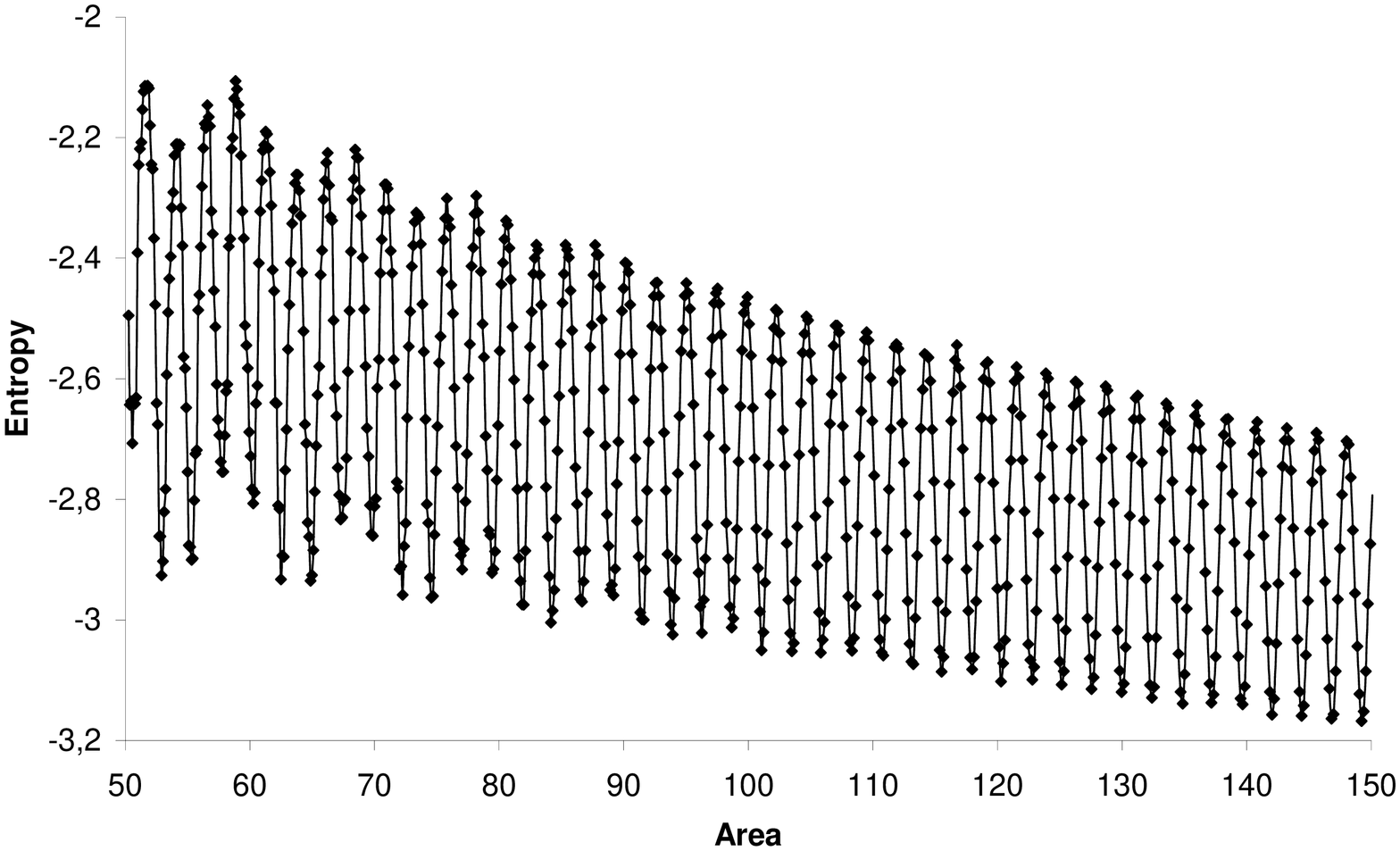}
  \includegraphics[angle=270,scale=.39]{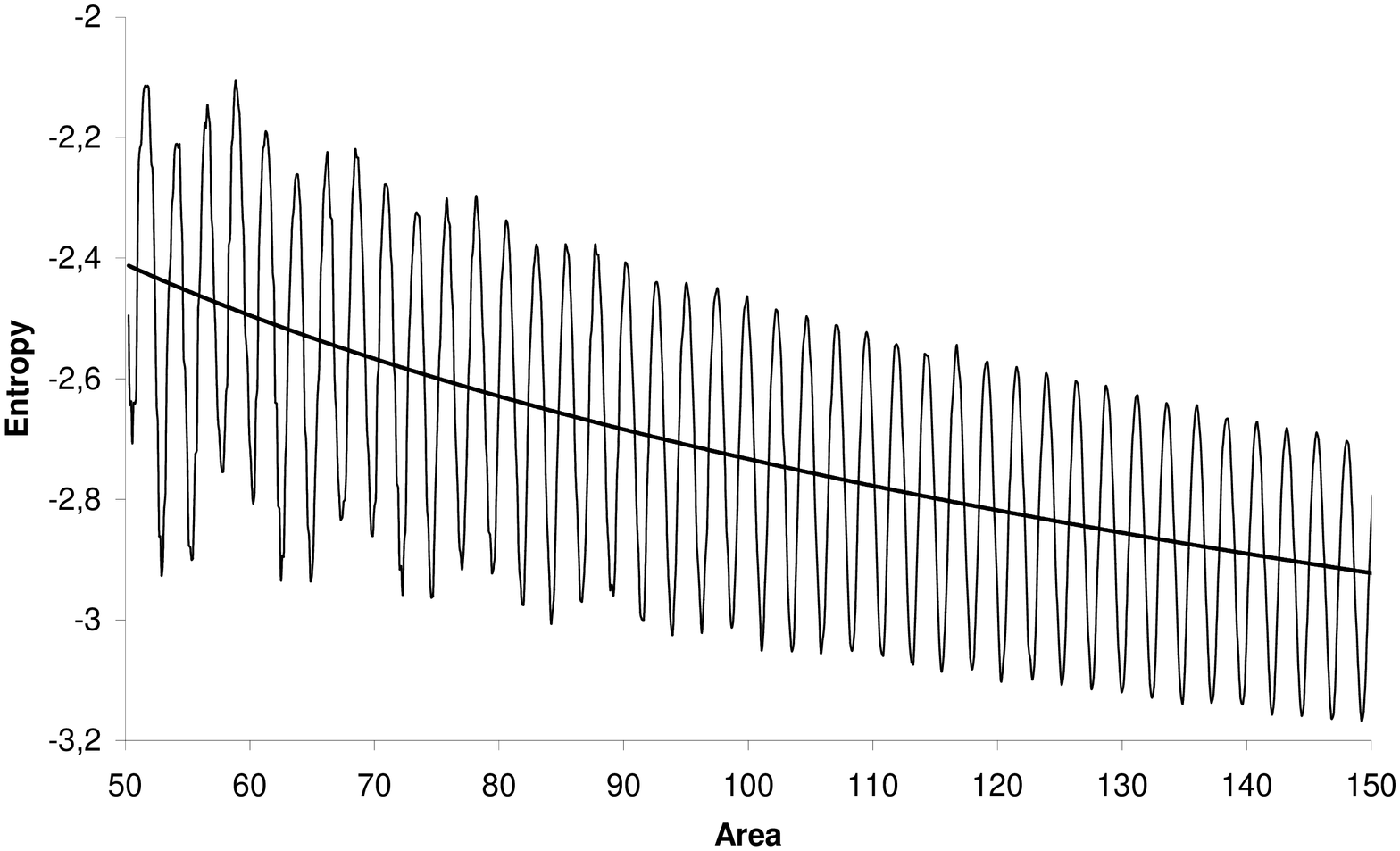} \caption{\label{fig:5}
  The curves of Fig.~\ref{fig:3}
are subtracted and the difference, an oscillatory function, shown
in the upper figure. The curve is approximated by a logarithm
curve in the lower figure.}
\end{figure}

When we include the projection constraint, the computation becomes
more involved and we are forced to consider a smaller range of
values for the area of the black holes. In Figure~\ref{fig:2}, we
plot both the entropy without the projection and with the
projection, keeping the same $\delta $. The first thing to note is
that for the computation with the constraint implemented, there
are some large oscillations in the number of states. Fitting a
straight line gives a slope of $0.237$. In order to reduce the
oscillations, we increased the size of $\delta $ to $\delta =2$.
The result is plotted in Figure~\ref{fig:3}. As can be seen the
oscillations are much smaller, and the result of implementing the
constraint is to shift the curve down (the slope is now 0.241). In
order to compare it with the expected behavior of
$S=A/4-(1/2)\ln{A}$, we subtracted both curves of
Figure~\ref{fig:3}, in the range $a_o=[50,160]$, and compared the
difference with a logarithmic function. The coefficient that gave
the best fit is equal to $-0.4831$ (See Figure~\ref{fig:5}). What
can we conclude from this? While it is true that the rapidly
oscillating function is far from the analytic curve, it is quite
interesting that the oscillatory function follows a logarithmic
curve with the ``right" coefficient. It is still a challenge to
understand the meaning of the oscillatory phase. Even when not
conclusive by any means, we can say that the counting of states is
consistent with a (n asymptotic) logarithmic correction with a
coefficient equal to (-1/2).

\section{Discussion}
\label{sec:6}

In this contribution we have considered the approach to the
quantum theory of gravity known as loop quantum gravity. We have
presented a brief introduction to the main ideas behind this
approach and have considered one of its main applications, black
hole entropy. We have discussed the main features in the approach
to black hole entropy, in particular in the implementation of the
isolated horizon boundary conditions to the quantum theory and how
this conditions tell us what states can be regarded as `black hole
states' that contribute to the entropy of the horizon. As we have
seen, the fact that there is an intrinsic discreteness in the
quantum horizon theory and that we are ignoring (tracing out) the
states in the bulk, is the reason why the entropy becomes finite.
It is sometimes believed that the fact that we do get an entropy
proportional to area is natural are not surprising, given that on
the horizon, the theory under consideration is a Chern-Simons
theory with punctures, and the entropy of a two dimensional theory
should be proportional to the total volume (area in this case). It
is important to stress that the result is not as trivial as it
sounds. To begin with, we do {\it not} have a given Chern-Simons
theory on the horizon, for any macro-state of a given area $a_o$,
there are many possible microstates that can be associated with
it. They do {\it not} all live on the same `Chern-Simons Hilbert
space'. The surface Hilbert space $\H_S$ is made out of the tensor
product of all possible Chern-Simons states compatible with the
constraints detailed in Sec.~ref{sec:5}, which belong to different
Chern-Simons states (characterized by, say, the total number of
punctures). That the total number of states compatible with the
(externally imposed) constraints is proportional to area is thus a
rather non-trivial result.

One might also wonder about the nature of the entropy one is
associating to the horizon. there has been some controversy about
the origin and location of the degrees of freedom responsible for
black hole entropy (see for instance \cite{MRT} for a recent
discussion). It has been argued that the degrees of freedom lie
behind the horizon, on the horizon and even on an asymptotic
region at infinity. What is then our viewpoint on this issue? The
viewpoint is that the IH boundary conditions implement in a
consistent manner an effective description, as horizon data, of
the degrees of freedom that might have formed the horizon. These
degrees of freedom, even if they are there in the physical
space-time, they are inaccessible to an external observer. The
only thing that this observer can `see' are the degrees of freedom
at the horizon, and these degrees of freedom are thus responsible
for the entropy associated to the horizon.

In the last part of this article, we have focussed our attention
on some recent results pertaining to the counting of states for
Planck size horizons. As we have shown, even when these black
holes lie outside the original domain of validity of the isolated
horizon formalism (tailored for large black holes), the counting
of such states has shed some light on such important issues as the
BI parameter, responsible for the asymptotic behavior, and the
first order, logarithmic, correction to entropy. We have also
found, furthermore, that there is a rich structure underlying the
area spectrum and the number of black hole states that could not
have been anticipated by only looking at the large area limit. In
particular, it has been found that the apparent periodicity in the
entropy {\it vs} area relation yields an approximate
`quantization´ of the entropy that makes contact with Bekenstein's
heuristic considerations \cite{CDF-2}, and is independent on the
choice of relevant states, and its associated counting. Details
will be published elsewhere \cite{CDF-4}.

\section*{Acknowledgments}

\noindent AC would like to thank the organizers of the NEB XII
International Conference in Nafplio, for the kind invitation to
deliver a talk on which this contribution is based. This work was
in part supported by CONACyT U47857-F, ESP2005-07714-C03-01 and
FIS2005-02761 (MEC) grants, by NSF PHY04-56913,  the Eberly
Research Funds of Penn State and by the AMC-FUMEC exchange
program. J.D. thanks MEC for a FPU fellowship.

\end{document}